\documentclass[aps,preprint,amsmath,amssymb,showpacs]{revtex4}

\usepackage{graphicx}
\begin{document}

\title{Anomalous $WW\gamma$ couplings in $\gamma p$ collision at the LHC}

\author{\.{I}. \c{S}ahin}
\email[]{inancsahin@karaelmas.edu.tr}
 \affiliation{Department of Physics, Zonguldak Karaelmas University, 67100 Zonguldak, Turkey}

\author{A. A. Billur}
\email[]{abillur@science.ankara.edu.tr} \affiliation{Department of
Physics, Cumhuriyet University, 58140 Sivas, Turkey}

\begin{abstract}
We examine the potential of $pp\to p\gamma p\to p W q X$ reaction to
probe anomalous $WW\gamma$ couplings at the LHC. We find 95\%
confidence level bounds on the anomalous coupling parameters with
various values of the integrated luminosity. We show that the
reaction $pp\to p\gamma p\to p W q X$ at the LHC highly improve the
current limits.
\end{abstract}


\maketitle

\section{Introduction}
Gauge boson self-interactions are consequences of the $SU_L(2)\times
U_Y(1)$ gauge structure of the standard model (SM). Measurements of
these couplings are crucial to test the non-Abelian structure of the
electroweak sector. Experimental results obtained from experiments
at CERN LEP and Fermilab Tevatron confirm the SM predictions.
Probing these couplings with a higher sensitivity can either lead to
additional confirmation of the SM or give some hint for new physics
beyond the SM.

\begin{figure}
\includegraphics{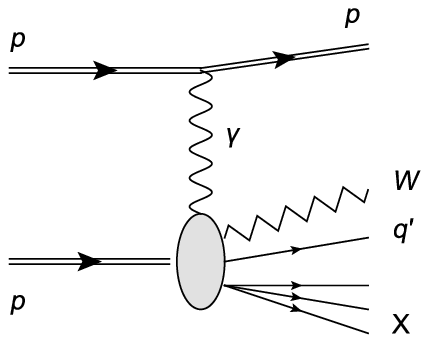}
\caption{Schematic diagram for the reaction $pp\to p\gamma p\to p W
q X $.\label{fig1}}
\end{figure}

In this work we have analyzed the anomalous $WW\gamma$ couplings via
single W boson production in $\gamma p$ collision at the LHC. A
quasi-real photon emitted from one proton beam can interact with the
other proton and produce W boson through deep inelastic scattering
(DIS). Since the emitted quasi-real photons have a low virtuality
they do not spoil the proton structure. Therefore the processes like
$pp\to p\gamma p\to p W q X $ can be studied at the LHC
(Fig.\ref{fig1}). Photon induced reactions in a hadron-hadron
collision were observed in the measurements of CDF collaboration
\cite{cdf1,cdf2,cdf3,cdf4}. For instance the reactions; $p \bar p\to
p \gamma \gamma \bar p\to p e^+ e^- \bar p$ \cite{cdf1,cdf4}, $p
\bar p\to p \gamma \gamma \bar p\to p\; \mu^+ \mu^- \bar p$
\cite{cdf3,cdf4}, $p \bar p\to p \gamma \bar p\to p\;
J/\psi\;(\psi(2S)) \bar p$ \cite{cdf3} were verified experimentally.
These results raise interest on the potential of LHC as a
photon-photon and photon-proton collider.

ATLAS and CMS collaborations have a program of forward physics with
extra detectors located at distances of 220m and 420m from the
interaction point \cite{royon,albrow}. The physics program of this
new instrumentation covers soft and hard diffraction, high energy
photon-induced interactions, low-x dynamics with forward jet
studies, large rapidity gaps between forward jets, and luminosity
monitoring
\cite{fd1,fd2,fd3,fd4,fd5,fd6,fd7,fd8,fd9,fd10,fd11,fd12,fd13,fd14,fd15,fd16,fd17,fd18}.
One of the main features of these forward detectors is to tag the
protons with some momentum fraction loss,
$\xi=(|\vec{p}|-|\vec{p}^{\,\,\prime}|)/|\vec{p}|$. Here $\vec{p}$
is the momentum of incoming proton and $\vec{p}^{\,\,\prime}$ is the
momentum of intact scattered proton. Complementary to proton-proton
interactions, forward detector equipment at the LHC allows to study
photon-photon and photon-proton interactions at energies higher than
at any existing collider.

New physics searches in photon-induced interactions at the LHC have
being discussed in the literature
\cite{fd11,fd12,fd13,lhc1,lhc2,lhc3,lhc4,lhc5,lhc6,lhc7,lhc8,lhc9}.
A detailed analysis of $WW\gamma$ couplings has been done in
\cite{lhc2} via the process $pp\to p \gamma \gamma p\to p W^+ W^-
p$. This process receives contributions both from anomalous
$WW\gamma$ and $WW\gamma\gamma$ couplings. On the other hand the
process $pp\to p\gamma p\to p W q X$ isolates $WW\gamma$ coupling
and gives us the opportunity to study $WW\gamma$ vertex independent
from $WW\gamma\gamma$. Therefore any signal which conflicts with the
SM predictions would be a convincing evidence for new physics
effects in $WW\gamma$.

\section{Cross sections and effective lagrangian}

We consider the following  subprocesses of the reaction $pp\to
p\gamma p\to p W q X$
\begin{eqnarray}
\label{subprocesses}
&&\text{(i)}\;\;\gamma u \to W^+ d
\;\;\;\;\;\;\;\;\;\;\;\;\text{(vi)}\;\;\gamma d \to W^- u \nonumber
\\ &&\text{(ii)}\;\;\gamma c \to W^+ s
\;\;\;\;\;\;\;\;\;\;\;\;\text{(vii)}\;\;\gamma s \to W^- c\nonumber
\\&&\text{(iii)}\;\;\gamma \bar d \to W^+ \bar u
\;\;\;\;\;\;\;\;\;\;\text{(viii)}\;\;\gamma b \to W^- t \\
&&\text{(iv)}\;\;\gamma \bar s \to W^+ \bar c
\;\;\;\;\;\;\;\;\;\;\;\text{(ix)}\;\;\gamma \bar u \to W^- \bar d
\nonumber \\&&\text{(v)}\;\;\gamma \bar b \to W^+ \bar t
\;\;\;\;\;\;\;\;\;\;\;\;\;\text{(x)}\;\;\gamma \bar c \to W^- \bar s
\nonumber
\end{eqnarray}
We neglect interactions between different family quarks since the
cross sections are suppressed  due to small off diagonal elements of
the CKM matrix.

Quasi-real photon which enters the subprocess is described by
equivalent photon approximation (EPA) \cite{budnev,Baur,fd18}.
Equivalent photon spectrum of virtuality $Q^2$ and energy $E_\gamma$
is given by
\begin{eqnarray}
\frac{dN_\gamma}{dE_{\gamma}dQ^{2}}=\frac{\alpha}{\pi}\frac{1}{E_{\gamma}Q^{2}}
[(1-\frac{E_{\gamma}}{E})
(1-\frac{Q^{2}_{min}}{Q^{2}})F_{E}+\frac{E^{2}_{\gamma}}{2E^{2}}F_{M}]
\end{eqnarray}
where
\begin{eqnarray}
&&Q^{2}_{min}=\frac{m^{2}_{p}E^{2}_{\gamma}}{E(E-E_{\gamma})},
\;\;\;\; F_{E}=\frac{4m^{2}_{p}G^{2}_{E}+Q^{2}G^{2}_{M}}
{4m^{2}_{p}+Q^{2}} \\
G^{2}_{E}=&&\frac{G^{2}_{M}}{\mu^{2}_{p}}=(1+\frac{Q^{2}}{Q^{2}_{0}})^{-4},
\;\;\; F_{M}=G^{2}_{M}, \;\;\; Q^{2}_{0}=0.71 \mbox{GeV}^{2}
\end{eqnarray}
Here E is the energy of the incoming proton beam and $m_{p}$ is the
mass of the proton. The magnetic moment of the proton is taken to be
$\mu^{2}_{p}=7.78$. $F_{E}$ and $F_{M}$ are functions of the
electric and magnetic form factors. The above EPA formula differs
from the pointlike electron positron case by taking care of the
electromagnetic form factors of the proton.

The cross section for the complete process $pp\to p\gamma p\to p W q
X$ can be obtained by integrating the cross section for the
subprocess $\gamma q \to W q^{\prime}$ over the photon and quark
spectra
\begin{eqnarray}
\sigma\left(pp\to p\gamma p\to p W q
X\right)=\int_{Q^{2}_{min}}^{Q^{2}_{max}} {dQ^{2}}\int_{x_{1\;
min}}^{x_{1\;max}} {dx_1 }\int_{x_{2\; min}}^{x_{2\;max}}
{dx_2} \nonumber \\
\times
\left(\frac{dN_\gamma}{dx_1dQ^{2}}\right)\left(\frac{dN_q}{dx_2}\right)\hat{\sigma}_{
\gamma q \to W q^{\prime}}(\hat s)\nonumber \\
=\int_{Q^{2}_{min}}^{Q^{2}_{max}}
{dQ^{2}}\int_{\frac{M_{inv}}{\sqrt{s}}}^{\sqrt{\xi_{max}}}
{dz\;2z}\int_{MAX(z^2,\;\xi_{min})}^{\xi_{max}}
{\frac{dx_1}{x_1}}\nonumber
\\ \times\left(\frac{dN_\gamma}{dx_1dQ^{2}}\right)N_q(z^2/x_1)\;\hat{\sigma}_{
\gamma q \to W q^{\prime}}(z^2s).
\end{eqnarray}
Here, $x_1=\frac{E_\gamma}{E}$ and $x_2$ is the momentum fraction of
the proton's momentum carried by the quark. Second integral in (5)
is obtained by transforming the differentials $dx_1dx_2$ into
$dzdx_1$ with a Jacobian determinant $2z/x_1$ where $z=\sqrt
{x_1x_2}\simeq\sqrt{\frac{\hat s}{s}}$. $M_{inv}$ is the total mass
of the final particles of the subprocess $\gamma q \to W
q^{\prime}$. $\frac{dN_q}{dx_2}$ is the quark distribution function
of the proton and $N_q(z^2/x_1)$ is $\frac{dN_q}{dx_2}$ evaluated at
$x_2=z^2/x_1$. At high energies greater than proton mass it is a
good approximation to write $\xi=\frac{E_\gamma}{E}=x_1$. The
virtuality of the quark is taken to be ${Q^\prime}^2={m_W}^2$ during
calculations. One should note that $Q^2$ and ${Q^\prime}^2$ refer to
different particles. In our calculations parton distribution
functions of Martin, Stirling, Thorne and Watt \cite{pdf} have been
used.

New physics contributions to $WW\gamma$ couplings can be
investigated in a model independent way by means of the effective
Lagrangian approach. The theoretical basis of such an approach rely
on the assumption that at higher energies beyond where the SM is
tested, there is a more fundamental theory which reduces to the SM
at lower energies. The SM is assumed to be an effective low-energy
theory in which heavy fields have been integrated out. Such a
procedure is quite general and independent of the new interactions
at the new physics energy scale. The charge and parity conserving
effective Lagrangian for $WW\gamma$ interaction can be written
following the papers \cite{gaemers,hagiwara}
\begin{eqnarray}
\label{lagrangian} \frac{iL}{g_{WW\gamma}}&&=g_{1}^{\gamma}
(W_{\mu\nu}^{\dagger}W^{\mu}A^{\nu}-
W^{\mu\nu}W_{\mu}^{\dagger}A_{\nu})+ \kappa
W_{\mu}^{\dagger}W_{\nu}A^{\mu\nu} +\frac{\lambda}
{M_{W}^{2}}W_{\rho\mu}^{\dagger} W_{\nu}^{\mu}A^{\nu\rho}
\end{eqnarray}
where
\begin{eqnarray}
g_{WW\gamma}=e \;\;, \;\;
V_{\mu\nu}=\partial_{\mu}V_{\nu}-\partial_{\nu}V_{\mu}\;\;,
\;\;V_\mu=W_\mu,A_\mu \nonumber
\end{eqnarray}
and dimensionless parameters $g_{1}^{\gamma}$, $\kappa$ and
$\lambda$ are related to the magnetic dipole and electric quadrupole
moments. The tree-level SM values for these parameters are
$g_{1}^{\gamma}=1$, $\kappa=1$ and $\lambda=0$. For on-shell
photons, $g_{1}^{\gamma}=1$ is fixed by electromagnetic gauge
invariance to its SM value at tree-level.

The vertex function for $W^{+}(p_1)W^{-}(p_2)A(p_3)$ generated from
the effective Lagrangian (\ref{lagrangian}) is given by
\begin{eqnarray}
\Gamma_{\mu\nu\rho}(p_{1},p_{2},p_{3})=&&e\left[g_{\mu\nu}
\left(p_{1}-p_{2}-\frac{\lambda}{M_{W}^{2}}[(p_{2}.p_{3})p_{1}
-(p_{1}.p_{3})p_{2}]\right)_{\rho} \right.\nonumber \\
&&\left.+g_{\mu\rho}\left(\kappa p_{3}-p_{1} + \frac{\lambda}
{M_{W}^{2}}[(p_{2}.p_{3})p_{1}
-(p_{1}.p_{2})p_{3}]\right)_{\nu}\right.\nonumber \\
&&\left.+g_{\nu\rho}\left(p_{2}-\kappa p_{3} -\frac{\lambda}
{M_{W}^{2}}[(p_{1}.p_{3})p_{2}
-(p_{1}.p_{2})p_{3}]\right)_{\mu} \right.\nonumber \\
&&\left.+\frac{\lambda}{M_{W}^{2}}(p_{2\mu}p_{3\nu}p_{1\rho}
-p_{3\mu}p_{1\nu}p_{2\rho})\right]
\end{eqnarray}
where  $p_1+p_2+p_3=0$.

During calculations we consider three different forward detector
acceptances; $0.0015 <\xi < 0.15$, $0.0015 <\xi < 0.5$ and $0.1 <\xi
< 0.5$. ATLAS Forward Physics (AFP) Collaboration proposed an
acceptance of $0.0015 <\xi < 0.15$ \cite{royon,albrow}. On the other
hand, CMS-TOTEM forward detector scenario spans $0.0015 <\xi < 0.5$
and $0.1 <\xi < 0.5$ \cite{avati,fd11}.

\begin{figure}
\includegraphics{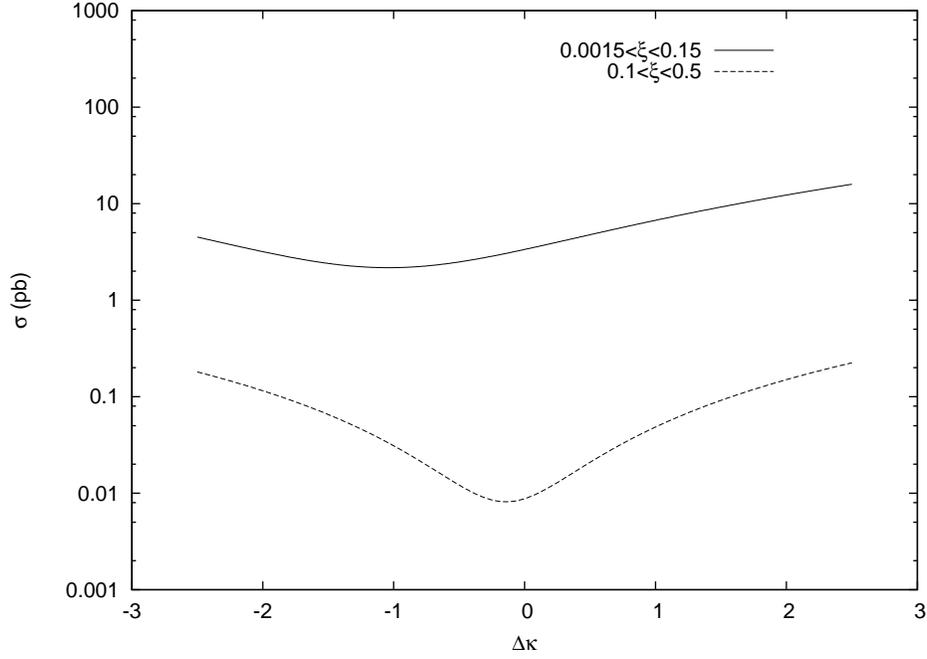}
\caption{The integrated total cross section of the process $pp\to
p\gamma p\to p W q X $ as a function of anomalous coupling $\Delta
\kappa=\kappa-1$ for two different forward detector acceptances
stated on the figure. We consider the sum of all subprocesses given
in equation (\ref{subprocesses}). The center of mass energy of the
proton-proton system is taken to be $\sqrt s=14$TeV.\label{fig2}}
\end{figure}

\begin{figure}
\includegraphics{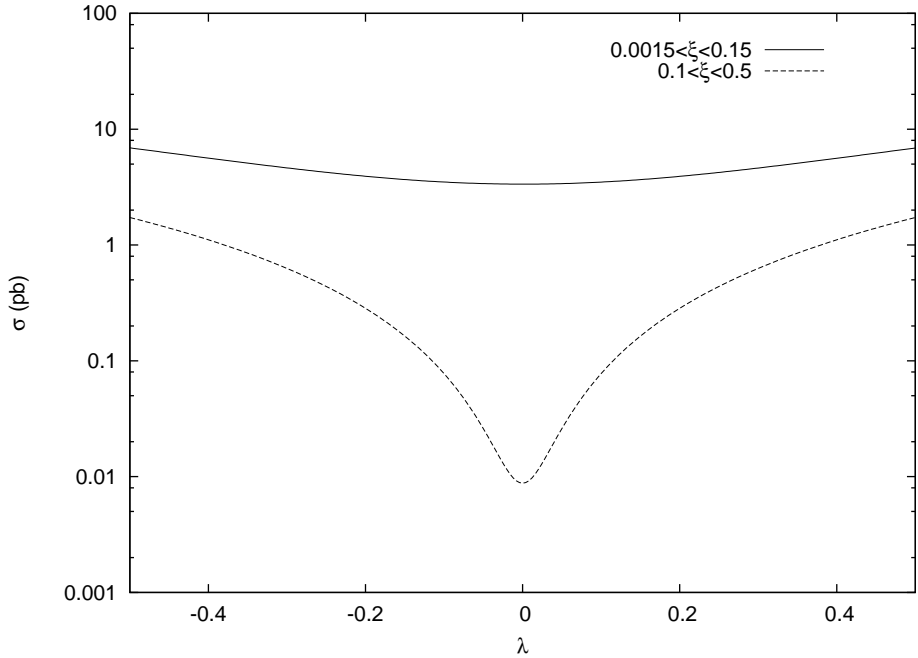}
\caption{The same as figure \ref{fig2} but for the coupling
$\lambda$. \label{fig3}}
\end{figure}

In Fig.\ref{fig2} and Fig.\ref{fig3} we plot the integrated total
cross section of the process $pp\to p\gamma p\to p W q X $ as a
function of anomalous couplings $\Delta \kappa=\kappa-1$ and
$\lambda$ for the acceptances $0.0015 <\xi < 0.15$ and $0.1 <\xi <
0.5$. We sum all the contributions from subprocesses given in
equation (\ref{subprocesses}). We do not plot the cross section for
the acceptance $0.0015 <\xi < 0.5$ since there is only a minor
difference between the curves for $0.0015 <\xi < 0.5$ and $0.0015
<\xi < 0.15$. In these figures we observe that although $0.1 <\xi <
0.5$ case gives small cross sections, deviations of the cross
sections from their SM values are large. This feature especially
remarkable for the cross section as a function of the coupling
$\lambda$.

\section{Sensitivity to anomalous couplings}

A detailed investigation of the anomalous couplings requires a
statistical analysis. To this purpose we have obtained 95\%
confidence level (C.L.) bounds on the anomalous coupling parameters
$\Delta \kappa=\kappa-1$ and $\lambda$ using one-parameter $\chi^2$
test. The $\chi^2$ function is given by,
\begin{eqnarray}
\chi^{2}=\left(\frac{\sigma_{SM}-\sigma(\Delta
\kappa,\lambda)}{\sigma_{SM} \,\, \delta}\right)^{2}
\end{eqnarray}
where $\delta=\frac{1}{\sqrt{N}}$ is the statistical error. The
expected number of events has been calculated considering the
leptonic decay channel of the W boson as the signal $N=0.9BR(W\to
\ell \nu)\sigma_{SM}L_{int}$, where $\ell=e$ or $\mu$ and 0.9 is the
survival probability factor \cite{fd12,fd11}. ATLAS and CMS have
central detectors with a pseudorapidity coverage $|\eta|<2.5$.
Therefore we place a cut of $|\eta|<2.5$ for electrons and muons
from decaying W and also for final quarks from subprocess $\gamma q
\to W q^{\prime}$.

In table \ref{tab1} and \ref{tab2}, we show 95\% C.L. sensitivity
bounds on the anomalous coupling parameters $\Delta \kappa$ and
$\lambda$ for various integrated luminosities and forward detector
acceptances of $0.0015<\xi<0.5$, $0.0015<\xi<0.15$ and
$0.1<\xi<0.5$. During statistical analysis only one of the anomalous
couplings is assumed to deviate from the SM at a time. We see from
the tables that bounds on $\Delta \kappa$ for $0.0015<\xi<0.5$ and
$0.0015<\xi<0.15$ cases are almost same. They are more than an order
of magnitude better than the bound obtained from $0.1<\xi<0.5$ case.
On the other hand, bounds on $\lambda$ are  more restrictive in
$0.1<\xi<0.5$ case with respect to $0.0015<\xi<0.5$ and
$0.0015<\xi<0.15$ cases. In table \ref{tab1} and \ref{tab2}, we
consider a center of mass energy of $\sqrt s=14$TeV for the
proton-proton system. But the LHC will not operate at $\sqrt s=14$
TeV before the year 2013. Therefore it is valuable to search its
sensitivity at $\sqrt s=7$ TeV. To this purpose we present table
\ref{tab3} where the sensitivity bounds are obtained at $\sqrt s=7$
TeV with an integrated luminosity of 1-2$fb^{-1}$.

The current bounds on anomalous $WW\gamma$ couplings are provided by
Fermilab Tevatron and CERN LEP. The most stringent bounds obtained
at the Tevatron are \cite{Abazov}
\begin{eqnarray}
\label{Tevatron}
-0.51<\Delta\kappa<0.51\;\;\;\;\;\;\;\;\;-0.12<\lambda<0.13
\end{eqnarray}
at 95\% C.L.. The combined fits of the four LEP experiments improves
the precision to \cite{Alcaraz}
\begin{eqnarray}
\label{LEP}
-0.098<\Delta\kappa<0.101\;\;\;\;\;\;\;\;\;-0.044<\lambda<0.047
\end{eqnarray}
at 95\% C.L.. Although the LEP bounds are more precise than the
bounds from the Tevatron, LEP results are obtained via the reactions
$e^-e^+\to W^-W^+$, $e^-e^+\to e\nu W$ and $e^-e^+\to \nu \bar \nu
\gamma$ where first two reactions receive contributions both from
$WW\gamma$ and $WWZ$ couplings. Therefore in general, limits given
in (\ref{LEP}) can not be regarded as a bound on pure $WW\gamma$
couplings.

\section{Conclusions}
LHC with a forward detector equipment gives us the opportunity to
study photon-photon and photon-proton interactions at energies
higher than at any existing collider. The reaction $pp\to p\gamma
p\to p W q X$ provides a rather clean channel compared to pure DIS
reactions due to absence of one of the incoming proton remnants.
Furthermore detection of the intact scattered protons in the forward
detectors allows us to reconstruct quasi-real photons momenta. This
may be useful in reconstructing the kinematics of the reaction.

The reaction $pp\to p\gamma p\to p W q X$ at the LHC with a
center-of-mass energy of 14 TeV probes anomalous $WW\gamma$
couplings with a better sensitivity than the LEP and Tevatron
experiments. Our limits also better than the limits obtained in
$pp\to p \gamma \gamma p\to p W^+ W^- p$ at the LHC \cite{lhc2}. We
also investigate the potential of the LHC with a center-of-mass
energy of 7 TeV. We deduce that the reaction $pp\to p\gamma p\to p W
q X$ with a center-of-mass energy of 7 TeV and an integrated
luminosity of 1$fb^{-1}$ probes anomalous $WW\gamma$ couplings with
a better sensitivity than Tevatron and with a comparable sensitivity
with respect to LEP.

A prominent advantage of the reaction $pp\to p\gamma p\to p W q X$
is that it isolates anomalous $WW\gamma$ couplings. It allows to
study $WW\gamma$ couplings independent from $WWZ$ as well as from
$WW\gamma\gamma$.

\pagebreak

\begin{table}
\caption{95\% C.L. sensitivity bounds of the coupling $\Delta\kappa$
for various forward detector acceptances and integrated LHC
luminosities. The center of mass energy of the proton-proton system
is taken to be $\sqrt s=14$TeV.\label{tab1}}
\begin{ruledtabular}
\begin{tabular}{ccccc}
Luminosity&$0.0015<\xi<0.5$ &$0.0015<\xi<0.15$ &$0.1<\xi<0.5$ \\
\hline
10$fb^{-1}$ &(-0.017 , 0.016)  &(-0.017 , 0.017) &(-0.428 , 0.146) \\
30$fb^{-1}$ &(-0.010 , 0.009) &(-0.010 , 0.010) &(-0.378 , 0.095) \\
50$fb^{-1}$ &(-0.007 , 0.007) &(-0.007 , 0.007) &(-0.360 , 0.078) \\
100$fb^{-1}$ &(-0.005 , 0.005) &(-0.005 , 0.005) &(-0.340 , 0.058) \\
200$fb^{-1}$ &(-0.004 , 0.004) &(-0.004 , 0.004) &(-0.325 , 0.043) \\
\end{tabular}
\end{ruledtabular}
\end{table}

\begin{table}
\caption{The same as table \ref{tab1} but for the coupling
$\lambda$. \label{tab2}}
\begin{ruledtabular}
\begin{tabular}{ccccc}
Luminosity&$0.0015<\xi<0.5$ &$0.0015<\xi<0.15$ &$0.1<\xi<0.5$ \\
\hline
10$fb^{-1}$ &(-0.043 , 0.044) &(-0.051 , 0.052) &(-0.017 , 0.017) \\
30$fb^{-1}$ &(-0.032 , 0.033) &(-0.039 , 0.040) &(-0.013 , 0.013) \\
50$fb^{-1}$ &(-0.028 , 0.029) &(-0.034 , 0.035) &(-0.011 , 0.011) \\
100$fb^{-1}$ &(-0.024 , 0.025) &(-0.029 , 0.030) &(-0.009 , 0.009) \\
200$fb^{-1}$ &(-0.020 , 0.021) &(-0.024 , 0.025) &(-0.008 , 0.008) \\
\end{tabular}
\end{ruledtabular}
\end{table}

\begin{table}
\caption{95\% C.L. sensitivity bounds of the couplings
$\Delta\kappa$ and $\lambda$ for various forward detector
acceptances and integrated LHC luminosities. The center of mass
energy of the proton-proton system is taken to be $\sqrt s=7$TeV.
\label{tab3}}
\begin{ruledtabular}
\begin{tabular}{cccccc}
&Luminosity&$0.0015<\xi<0.5$ &$0.0015<\xi<0.15$ &$0.1<\xi<0.5$ \\
\hline
Limits on $\Delta\kappa$:& & & \\
&1$fb^{-1}$ &(-0.081, 0.075) &(-0.086, 0.080) &(-1.084, 0.305) \\
&2$fb^{-1}$ &(-0.057, 0.054) &(-0.060, 0.057) &(-1.010, 0.231) \\
Limits on $\lambda$:& & & \\
&1$fb^{-1}$ &(-0.144, 0.145) &(-0.176, 0.178) &(-0.078, 0.078) \\
&2$fb^{-1}$ &(-0.121, 0.122) &(-0.148, 0.150) &(-0.066, 0.066) \\
\end{tabular}
\end{ruledtabular}
\end{table}

\end{document}